# Interacting Vector-Spinor and Nilpotent Supersymmetry

Hitoshi NISHINO[1]) and Subhash RAJPOOT[2])

*Department of Physics & Astronomy*
*California State University*
*1250 Bellflower Boulevard*
*Long Beach, CA 90840*

## Abstract

We formulate an interacting theory of a vector-spinor field that gauges anti-commuting spinor charges $\{Q_\alpha{}^I, Q_\beta{}^J\} = 0$ in arbitrary space-time dimensions. The field content of the system is $(\psi_\mu{}^{\alpha I}, \chi^{\alpha IJ}, A_\mu{}^I)$, where $\psi_\mu{}^{\alpha I}$ is a vector-spinor in the adjoint representation of an arbitrary gauge group, and $A_\mu{}^I$ is its gauge field, while $\chi^{\alpha IJ}$ is an extra spinor with antisymmetric adjoint indices $IJ$. Amazingly, the consistency of the vector-spinor field equation is maintained, despite its non-trivial interactions.

---

[1]) E-Mail: hnishino@csulb.edu

[2]) E-Mail: rajpoot@csulb.edu

## 1. Introduction

It has been common wisdom that there exist no consistent non-trivial interactions for spin 3/2 (vector-spinor) field [1][2] other than in supergravity theory with local supersymmetry [3][4][5][6][7][8]. However, it is not clear, if this statement applies also to the case of nilpotent anti-commuting spinor charges $Q_\alpha{}^I$ satisfying

$$\{Q_\alpha{}^I, Q_\beta{}^J\} = 0 ~, \tag{1.1a}$$

$$[T^I, Q_\alpha{}^J] = f^{IJK} Q_\alpha{}^K ~, \tag{1.1b}$$

where the $T^I$'s are the usual antihermitian generators of an arbitrary non-Abelian gauge group $G$ with the commutator

$$[T^I, T^J] = f^{IJK} T^K ~, \tag{1.2}$$

with the adjoint indices $I$, $J$, $K$ = 1, 2, $\cdots$, dim $G$. Due to the absence of the r.h.s. of (1.1a), we call these spinor charges 'nilpotent supersymmetry'. The gauging of nilpotent spinor charges is nothing new by itself. A typical example is the gauging [9] of BRST symmetry [10][11]. However, the conventional BRST charges have no additional index, while our spinor charge in (1.1) carries the spinorial index $\alpha$ as well as the adjoint index $I$ of the non-Abelian gauge group $G$.

In this paper, we present for the first time non-trivial consistent interactions for the vector-spinor field $\psi_\mu{}^{\alpha I}$ gauging the nilpotent supersymmetry $Q_\alpha{}^I$. We will show that our theory has no problem at the classical level, such as the absence of negative energy ghosts for the vector spinor with the standard kinetic term at the bilinear order, and the absence of inconsistent interactions. Our formulation is similar to the non-Abelian tensor formulation in our recent paper [12], inspired by the generalized dimensional reduction by Scherk-Schwarz [13], in which we constructed the whole multiplet $(B_{\mu\nu}{}^I, C_\mu{}^{IJ}, K^{IJK}, A_\mu{}^I)$ with the consistent field strength for the non-Abelian tensor $B_{\mu\nu}{}^I$. In our present formulation, we have the field content $(\psi_\mu{}^{\alpha I}, \chi^{\alpha IJ}, A_\mu{}^I)$,[3] where the vector-spinor $\psi_\mu{}^I$ carries the adjoint index $I$, while the extra spinor $\chi^{IJ}$ carries antisymmetric indices $IJ$. We confirm the two main important ingredients: First, we can define the field strength of $\psi_\mu{}^{\alpha I}$ which is invariant under the spinor symmetry $Q_\alpha{}^I$ and covariant under the usual non-Abelian gauge symmetry. Second, we have the consistency of the $\psi_\mu{}^I$-field equation, which is usually very difficult to accomplish with non-trivial interactions without supergravity [3][4][5]. We carry out these objectives in arbitrary space-time dimensions.

---

[3] We sometimes omit the spinorial indices $\alpha$, $\beta$, $\cdots$ on fermions.

## 2. Lagrangian and Transformation Rules

Our field content is $(\psi_\mu{}^{\alpha I}, \chi^{\alpha IJ}, A_\mu{}^I)$, where $\psi_\mu{}^{\alpha I}$ is the Majorana spinor with the vectorial index gauging the spinor charge $Q_\alpha{}^I$ in (1.1). Our space-time dimensions $D$ is arbitrary throughout this paper with the signature diag.$(-,+,+,\cdots,+)$, in which the Majorana spinors $\psi_\mu{}^{\alpha I}$ and $\chi^{\alpha IJ}$ can be always defined [14].

Our total action $I \equiv I_1 + I_2 + I_3$ has the three lagrangians $\mathcal{L}_i$ $(i = 1, 2, 3)$ defined by[4]

$$I_1 \equiv \int d^D x\, \mathcal{L}_1 \equiv \int d^D x \left[ +\tfrac{1}{4} a_0^{-1} f^{IJK} (\overline{L}_\mu{}^{IJ} \gamma^{\mu\nu\rho} \mathcal{R}_{\nu\rho}{}^K) \right] \ , \tag{2.1a}$$

$$I_2 \equiv \int d^D x\, \mathcal{L}_2 \equiv \int d^D x \left[ +\tfrac{1}{2} P^{IJ,KL} (\overline{\chi}^{IJ} \gamma^\mu L_\mu{}^{KL}) \right] \ , \tag{2.1b}$$

$$I_3 \equiv \int d^D x\, \mathcal{L}_3 \equiv \int d^D x \left[ -\tfrac{1}{4} (F_{\mu\nu}{}^I)^2 \right] \ , \tag{2.1c}$$

where $f^{IJK}$ is the structure constant of the gauge group $\forall G$, while the real positive constant $a_0 > 0$ and the $P$'s are defined by

$$f^{IJK} f^{JKL} \equiv a_0\, \delta^{IL} \ , \quad P^{IJ,KL} \equiv \delta^{[I|K} \delta^{|J]L} - Q^{IJ,KL} \ , \quad Q^{IJ,KL} \equiv a_0^{-1} f^{IJM} f^{MKL} \ . \tag{2.2}$$

Here the $P$'s and $Q$'s are projectors satisfying

$$P^{IJ,KL} + Q^{IJ,KL} = \delta^{[I|K} \delta^{|J]L} \ , \quad P^{IJ,KL} Q^{KL,MN} = 0 \ , \tag{2.3a}$$

$$P^{IJ,KL} P^{KL,MN} = P^{IJ,MN} \ , \quad Q^{IJ,KL} Q^{KL,MN} = Q^{IJ,MN} \ . \tag{2.3b}$$

The $\mathcal{R}$'s and $L$'s are the field strengths of $\psi$ and $\chi$ defined in a peculiar way by

$$\mathcal{R}_{\mu\nu}{}^I \equiv 2 D_{[\mu} \psi_{\nu]}{}^I + F_{\mu\nu}{}^J \chi^{IJ} \ , \tag{2.4a}$$

$$L_\mu{}^{IJ} \equiv D_\mu \chi^{IJ} + f^{IJK} \psi_\mu{}^K \ , \tag{2.4b}$$

where the $D$'s is the standard non-Abelian covariant derivative, e.g., $D_\mu \psi_\nu{}^I \equiv \partial_\mu \psi_\nu{}^I + f^{IJK} A_\mu{}^J \psi_\nu{}^K$, etc. The second term in (2.4a) is understood as a Chern-Simons term, while the $\psi$-linear term in (2.4b) is the covariantization of its first term under the nilpotent supersymmetry as will be seen (2.6) below. The $\psi$-linear term in the $L$'s explains how the lagrangian $\mathcal{L}_1$ contains the standard kinetic term for $\psi_\mu$.

As is clear from (2.3), the $P$'s and $Q$'s are nothing but the projection operators. Namely, the $g(g-1)/2$-dimensional indices $IJ$ are projected out by $P$'s and $Q$'s respectively into $g(g-3)/2$ and $g$-dimensional subspaces [12]. It is now clear that the $P$-projector in $\mathcal{L}_2$ projects out the original $g(g-1)/2$ components in $\chi^{IJ}$ into $g(g-3)/2$ components.

---

[4] We use always superscripts for $I, J, \cdots$ due to the positive definite signature for the compact gauge group manifold.

Only those $g(g-3)/2$ components have the kinetic energy, while the remaining $g$ components are just *gauge* degrees of freedom, as will be clarified by the nilpotent supersymmetry (2.6) below.

Relevantly, the field strengths $\mathcal{R}$ and $L$ satisfy their Bianchi identities

$$D_{[\mu}\mathcal{R}_{\nu\rho]}{}^I \equiv +F_{[\mu\nu}{}^J L_{\rho]}{}^{IJ} \ . \tag{2.5a}$$

$$D_{[\mu}L_{\nu]}{}^{IJ} \equiv +\tfrac{1}{2}f^{IJK}\mathcal{R}_{\mu\nu}{}^K - \tfrac{3}{2}f^{[IJ|K}F_{\mu\nu}{}^L \chi^{K|L]} \ . \tag{2.5b}$$

Note that all the indices $I$, $J$, $L$ in the last term are totally antisymmetrized.

Our nilpotent supersymmetry transformation rule is

$$\delta_\epsilon \psi_\mu{}^I = D_\mu \epsilon^I \ , \qquad \delta_\epsilon A_\mu{}^I = 0 \ , \tag{2.6a}$$

$$\delta_\epsilon \chi^{IJ} = -f^{IJK}\epsilon^K = -Q^{IJ,KL}f^{KLM}\epsilon^M \ , \tag{2.6b}$$

where $\epsilon^{\alpha I}$ is Majorana spinor parameter for $Q_\alpha{}^I$. This transformation rule is fixed by studying the parallel case for bosonic non-Abelian tensor [12]. Relevantly, the field strengths are invariant:

$$\delta_\epsilon \mathcal{R}_{\mu\nu}{}^I = 0 \ , \qquad \delta_\epsilon L_\mu{}^{IJ} = 0 \ . \tag{2.7}$$

From these relations, the projector $P$ leaves the lagrangian $\mathcal{L}_2$ invariant under $\delta_\epsilon$, because $\delta_\epsilon(P^{IJ,KL}\chi^{KL}) = 0$, even though $\delta_\epsilon \chi^{IJ} \neq 0$. Therefore our actions $I_1$, $I_2$ and $I_3$ are separately invariant under $\delta_\epsilon$. It is also clear that the commutator $[\delta_{\epsilon_1}, \delta_{\epsilon_2}]$ is vanishing consistently with the r.h.s. of (1.1).

As has been mentioned, the components in $\chi^{IJ}$ projected out by the $Q$'s are just gauge degrees of freedom, and this can be seen in (2.6b). In order to confirm this point more rigorously, we introduce a new parameter $\Lambda^{IJ}$ defined by

$$\Lambda^{IJ} \equiv +Q^{IJ,KL}f^{KLM}\epsilon^M = f^{IJK}\epsilon^K \ , \tag{2.8}$$

so that

$$\epsilon^I = a_0^{-1}f^{IJK}\Lambda^{JK} \ , \qquad P^{IJ,KL}\Lambda^{KL} \equiv 0 \ , \qquad Q^{IJ,KL}\Lambda^{KL} \equiv \Lambda^{IJ} \ . \tag{2.9}$$

In other words, $\Lambda^{IJ}$ has components only in the direction of the $Q$-projector. Accordingly, (2.6a) and (2.6b) are rewritten as the extra $\delta_\Lambda$-symmetry of the action:

$$\delta_\Lambda \psi_\mu{}^I = D_\mu \left(a_0^{-1}f^{IJK}\Lambda^{JK}\right) \ , \tag{2.10a}$$

$$\delta_\Lambda \chi^{IJ} = -\Lambda^{IJ} = -Q^{IJ,KL}\Lambda^{KL} \ . \tag{2.10b}$$

Eq. (2.10b) implies explicitly that the components in $\chi^{IJ}$ projected out by the $Q$'s are purely gauge degrees of freedom, while inducing the usual nilpotent supersymmetry (2.10a) for $\psi_\mu{}^I$. Note that the invariance $\delta_\Lambda I = 0$ is not only for the free kinetic term in $\mathcal{L}_2$, but to all orders of interactions.

This symmetry is very important, because the unphysical $Q$-direction of the $\chi^{IJ}$-field is gauged away, and it never enters higher-order interactions with physical fields. Since these unphysical components lack their kinetic terms in $\mathcal{L}_2$, if they entered higher-order interactions with physical fields, their field equations would yield undesirable constraints upon physical fields. Thanks to the all-order symmetry (2.10), the unphysical components in $\chi^{IJ}$ are completely gauged away to all orders, and excluded from interactions with physical fields.

Our action has also the local non-Abelian gauge symmetry for the gauge group $^\forall G$:

$$\delta_\alpha A_\mu{}^I = D_\mu \alpha^I \ , \tag{2.11a}$$

$$\delta_\alpha \psi_\mu{}^I = -f^{IJK} \alpha^J \psi_\mu{}^K \ , \tag{2.11b}$$

$$\delta_\alpha \chi^{IJ} = -2 f^{[I|KL} \alpha^K \chi^{L|J]} \ . \tag{2.11c}$$

Accordingly, we have also the relationships

$$\delta_\alpha \mathcal{R}_{\mu\nu}{}^I = -f^{IJK} \alpha^J \mathcal{R}_{\mu\nu}{}^K \ , \tag{2.12a}$$

$$\delta_\alpha L_\mu{}^{IJ} = -2 f^{[I|KL} \alpha^K L_\mu{}^{L|J]} \ . \tag{2.12b}$$

These are homogeneous local non-Abelian transformations, we have the invariance $\delta_\alpha I = 0$. We can also confirm the closures of all the commutators in (1.1) and (1.2).

The field equations for the $\psi_\mu$, $\chi$ and $A_\mu$-fields are[5])

$$\frac{\delta \mathcal{L}}{\delta \overline{\psi}_\mu{}^I} = +\tfrac{1}{2} \gamma^{\mu\rho\sigma} \mathcal{R}_{\rho\sigma}{}^I - \tfrac{1}{4} \gamma^{\mu\rho\sigma} \chi^{IJ} F_{\rho\sigma}{}^J + \tfrac{1}{4} Q^{IJ,KL} \gamma^{\mu\rho\sigma} \chi^{KL} F_{\rho\sigma}{}^J \doteq 0 \ , \tag{2.13a}$$

$$\frac{\delta \mathcal{L}}{\delta \overline{\chi}^{IJ}} = + P^{IJ,KL} \gamma^\mu L_\mu{}^{KL} - \tfrac{1}{4} a_0^{-1} f^{IJK} \gamma^{\mu\rho\sigma} L_\mu{}^{KL} F_{\rho\sigma}{}^L$$
$$+ \tfrac{1}{4} a_0^{-1} f^{[I|KL} \gamma^{\mu\rho\sigma} L_\mu{}^{KL} F_{\rho\sigma}{}^{|J]} \doteq 0 \ , \tag{2.13b}$$

$$\frac{\delta \mathcal{L}}{\delta A_\mu{}^I} = - D_\nu F^{\mu\nu\, I} - \tfrac{1}{4} Q^{IJ,KL} (\overline{\chi}^{KL} \gamma^{\mu\rho\sigma} \mathcal{R}_{\rho\sigma}{}^J) + \tfrac{1}{2} Q^{IJ,KL} (\overline{\psi}_\rho{}^J \gamma^{\mu\rho\sigma} L_\sigma{}^{KL})$$
$$- \tfrac{1}{2} a_0^{-1} f^{JKL} D_\nu (\overline{\chi}^{IL} \gamma^{\mu\nu\rho} L_\rho{}^{JK}) - f^{IJK} (\overline{\chi}^{JL} \gamma^\mu \chi^{KL})$$
$$+ \tfrac{1}{2} f^{KLJ} Q^{JI,MN} (\overline{\chi}^{MN} \gamma^\mu \chi^{KL}) \doteq 0 \ . \tag{2.13c}$$

---

[5]) In this paper we use the symbol $\doteq$ for a field equation. We also use the symbol $\stackrel{?}{=}$ for an equation under question.

The linear terms of these equations are the kinetic terms of the vector-spinor $\psi_\mu{}^I$ and the physical $g(g-3)/2$-components of $\chi^{IJ}$. In particular, due to the relationship $P^{IJ,KL} f^{KLM} = 0$, the $\psi$-linear term in the first term in (2.13b) does not contribute, and therefore at the linear order there is no mixture of $\psi$ with the $\chi$-field equation. Additionally, as the linear terms in (2.13a) and (2.13b) show, both $\psi$ and $\chi$ are massless.

## 3. Consistency of Vector-Spinor Field Equation

There are a few remarks on our system: First, we see that the $F\chi$-term in the definition of (2.4a) is important for the invariance of $\mathcal{R}$. If this term *were not* in $\mathcal{R}$, then there would be a term proportional to $F_{\mu\nu}{}^{IJ} \epsilon^J$ in $\delta_\epsilon \mathcal{R}_{\mu\nu}{}^I$, spoiling the invariance of $\mathcal{R}$. Second, this $F\chi$-term and the $\psi$-linear term in (2.4b) are the analog of the generalized dimensional reduction by Scherk-Schwarz [13]. Namely, analogously to the bosonic non-Abelian tensor formulation in [12], we need the extra spinor field $\chi^{IJ}$ with extra $IJ$ indices, so that the numbers of space-time indices and the adjoint indices add up to three for both $\psi_\mu{}^I$ and $\chi^{IJ}$.

Third, these peculiar forms of field strengths are also related to the consistency of the $\psi$-field equation (2.13a). Consider the divergence of the $\psi$-field equation $D_\mu(\delta\mathcal{L}/\delta\psi_\mu{}^I) \stackrel{?}{=} 0$. Since the inside of the parentheses vanishes as the $\psi$-field equation, its divergence should also vanish. This problem with a vector-spinor has been known to be very difficult to solve for a long time [15], unless we have local supersymmetry [3][4]. Fortunately, in our system, we have the $\delta_\epsilon$-invariance of the action $I$, and therefore we have the identity

$$D_\mu\left(\frac{\delta\mathcal{L}}{\delta\psi_\mu{}^I}\right) + f^{IJK}\left(\frac{\delta\mathcal{L}}{\delta\chi^{JK}}\right) \equiv 0 \ . \tag{3.1}$$

Note that this is an *identity* without any use of field equation, and is nothing but a rewriting of $\delta_\epsilon I = 0$. Eq. (3.1) immediately implies that by the use of the $\chi$-field equation, the above divergence of the $\psi$-field equation vanishes, as desired. Note that this is closely related to the $F\chi$-term or $\psi$-linear term in the $\mathcal{R}$ and $L$-field strengths, because without these terms our action is not invariant, and consequently we will fail to get (3.1). We stress that our system is the first system other than supergravity [3][4][5] that maintains consistent and non-trivial interactions for a propagating vector spinor.

We can also confirm (3.1) directly using the field equations (2.13). To this end, the important relationships we need are the Bianchi identities (2.5), and a useful lemma $Q^{IJ,KL} f^{KLM} = f^{IJM}$.

Even though the invariance $\delta_\alpha I = 0$ implies also the consistency of the $A_\mu$-field equation, it is a good verification of our field equations (2.13) to confirm the consistency $D_\mu(\delta\mathcal{L}/\delta A_\mu{}^I) \overset{?}{=} 0$ explicitly. In fact, we can arrange all the terms in $D_\mu(\delta\mathcal{L}/\delta A_\mu{}^I)$ into two groups, one vanishing upon the use of the $\psi$-field equation (2.13a), and another upon the $\chi$-field equation (2.13b), in agreement with $\delta_\alpha I = 0$:

$$D_\mu\left(\frac{\delta\mathcal{L}}{\delta A_\mu{}^I}\right) \equiv -f^{IJK}\overline{\psi}_\mu{}^J\left(\frac{\delta\mathcal{L}}{\delta\overline{\psi}_\mu{}^K}\right) + 2f^{I[J|L}\overline{\chi}^{L|K]}\left(\frac{\delta\mathcal{L}}{\delta\overline{\chi}^{JK}}\right) \ . \tag{3.2}$$

Note that this is an identity, and is *not* a result of field equations. Since each term here vanishes upon the $\psi$ and $\chi$-field equations, (3.2) confirms the consistency of our $A_\mu$-field equation (2.13c).

There is one subtlety in our system to be mentioned. Some readers may wonder, if the $\chi$-field is just a compensator for the nilpotent supersymmetry $Q_\alpha{}^I$, and therefore $\psi_\mu{}^I$ is not really its gauge field. There are two points to be mentioned to clarify this.

First, the problem of 'fake' gauge field arises, when its kinetic term is absent. For example, if we define an $U(1)$ 'gauge field' by $A_\mu \equiv \partial_\mu\varphi$ with the real scalar *compensator* $\varphi$, and a complex scalar $\phi$ carrying the $U(1)$ charge, their infinitesimal transformations are

$$\delta_\alpha A_\mu = +\partial_\mu\alpha \ , \quad \delta_\alpha\varphi = +\alpha \ , \quad \delta_\alpha\phi = -i\alpha\phi \ , \quad \delta_\alpha\phi^* = +i\alpha\phi^* \ . \tag{3.3}$$

Now with the covariant derivative is $D_\mu\phi \equiv \partial_\mu\phi - ieA_\mu\phi \equiv \partial_\mu\phi - ie(\partial_\mu\varphi)\phi$, the kinetic term of $\phi$ is

$$I_\phi \equiv \int d^D x \ [-(D_\mu\phi)^*(D^\mu\phi)] \ . \tag{3.4}$$

Even though the action has the local $U(1)$ invariance $\delta_\alpha I_\phi = 0$, the $A_\mu$ is a 'fake' gauge field with no kinetic term. Another way to see this is the field-redefinitions

$$\widetilde{\phi} \equiv e^{-ie\varphi}\phi \quad \Longrightarrow \quad D_\mu\phi = e^{ie\varphi}\partial_\mu\widetilde{\phi} \ , \tag{3.5}$$

so that $\varphi$ is completely absent from the action $I_\phi$. Even if we try to put a formal $A_\mu$-kinetic term $\mathcal{L}_{F^2} \equiv -(F_{\mu\nu})^2/4$, it is identically zero, because $A_\mu$ is pure-gauge. However, this problem of fake gauge field does not arise in our system. In fact, our $\psi$-kinetic term (2.1b) is non-trivial and not identically vanishing, because *not all* the components in the $\psi$'s are pure-gauge components, as (2.6) shows. This is the considerable difference from the compensator system presented in (3.3).

Second, an analogous example exists for the Lorentz connection $\omega_\mu{}^{rs}$. In the conventional formulation of gravity with local Lorentz symmetry $\mathcal{M}_{rs}$, we can regard $\omega_\mu{}^{rs}(e) \equiv (C_\mu{}^{rs} - C_\mu{}^{sr} - C^{rs}{}_\mu)/2$ as a function of the anholonomy coefficient $C_{\mu\nu}{}^m \equiv \partial_\mu e_\nu{}^m - \partial_\nu e_\mu{}^m$. Since the antisymmetric part of the vierbein $e_{\mu m}$ transforms as a 'compensator' for the local Lorentz symmetry, $\omega(e)$ has no degree of freedom, and it is a 'fake' gauge field.[6)]

However, our system of nilpotent supersymmetry above has an important difference from the Lorentz connection $\omega_\mu{}^{rs}(e)$, because $\psi_\mu{}^{\alpha I}$ as the gauge field for $Q_\alpha{}^I$ has its own kinetic term. This is analogous to a formulation, in which the contorsion tensor part $K_\mu{}^{rs} \equiv (T_\mu{}^{rs} - T_\mu{}^{sr} - T^{rs}{}_\mu)/2$ in the Lorentz connection

$$\omega_\mu{}^{rs} = \omega_\mu{}^{rs}(e) + K_\mu{}^{rs} \quad , \tag{3.6}$$

has its own physical degrees of freedom as a 'propagating torsion'. This is also equivalent to regarding $\omega_\mu{}^{rs}$ as an independent gauge field for the generator $\mathcal{M}_{rs}$ with its proper kinetic term, such as the curvature tensor squared $\approx R_{\mu\nu rs}(\omega)R^{\mu\nu rs}(\omega)$, *etc.*, sometimes known as Poincaré gauge theory [16]. The two formulations: the first one for propagating contorsion $K$, and the second one for propagating Lorentz connection $\omega$, are equivalent to each other by the equality (3.6). The first term in (3.6) transforms as a gradient, while $K$ transforms as a tensor under $\mathcal{M}_{rs}$. We can reformulate $\omega$ in [16] in terms of propagating torsion $K$ instead of $\omega$ itself, by shifting the anholonomy part $\omega(e)$ by (3.6). In other words, both sides of (3.3) have the same degrees of freedom, and we can interpret $\omega$ as an independent field variable. Mimicking this analogy [16], we can perform a field redefinition similar to (3.3) absorbing the 'compensator' $\chi^{IJ}$ based on (2.10), but we still have the kinetic term for the leading part of $\psi$ showing non-trivial features for our system.

From these considerations, it is clear that our system is non-trivial for the vector spinor $\psi_\mu{}^{\alpha I}$ with consistent interactions, whether we regard this as a gauge field of the nilpotent supersymmetry $Q_\alpha{}^I$, or subtract partial components in the gradient part in terms of the compensator $\chi$.

## 4. Concluding Remarks

In this paper, we have presented a new formulation for the vector-spinor gauge field $\psi_\mu{}^I$ with non-trivial consistent interactions in $\forall D$ for non-Abelian gauge group $\forall G$. Even though there is a similarity between our nilpotent spinor charge $Q_\alpha{}^I$ and the BRST charges [10][9], the former has spinorial and adjoint indices that the latter lacks. We have also confirmed the consistency of the $\psi$-field equation despite non-trivial interactions, both with the gauge field and the extra spinor $\chi^{IJ}$. To our knowledge, our formulation is the only one other than supergravity [3][4][5], that has consistent interactions for a *physical* vector-spinor field in arbitrary space-time dimensions.

---

[6)] It is usually called a 'composite' gauge field, but such terminology is not the issue here.

Compared with our recent bosonic non-Abelian tensor formulation [12], there are similarities as well as differences. One of the similarities is that we need both $\psi_\mu{}^I$ and $\chi^{IJ}$ in order to have the invariant field strength $\mathcal{R}_{\mu\nu}{}^I$. Another similarity is the structure of the indices, *i.e.*, the total number of spacial and adjoint indices should be maintained. One of the differences is that the kinetic term of the vector-spinor should have a peculiar structure of combination of $\mathcal{L}$ and $\mathcal{R}$-field strength, where the former carries the bare $\psi$-term. Since a fermionic kinetic term has the structure of (Potential)$\times$(Field Strength) instead of (Field Strength)$^2$, this difference is inevitable. Another difference is that all the fermionic fields remain massless, contrary to the massive non-Abelian tensor in the bosonic case [12].

Note that there is no upper limit for the number of gravitini in our formulation, because the dimensionality $\dim G$ is not bounded from above. In this sense, our theory is similar to so-called '$\aleph_0$-extended supersymmetric Chern-Simons theory' [17] or '$\aleph_0$-hypergravity' [18] in 3D, where non-dynamical gravitini carry the adjoint indices of an arbitrary gauge groups. However, in our present formulation, the gravitini are always *physical* with propagating degrees of freedom. Another difference is that our formulation is valid in $^\forall D$, while those in [17][18] are valid only in 3D. Even though our algebra (1.1a) seems to be too simple to have solid physical content at first glance, we have seen that it has such intricate structures with non-trivial interactions with a non-Abelian vector-spinor field in $^\forall D$.

We are grateful to W. Siegel for important discussions. This work is supported in part by NSF Grant # 0308246.

# References

[1] A.S. Wightman in *'Invariant Wave Equations'*, Proceedings of the Ettore Majorana International School of Mathematical Physics, Erice, 1977, *eds.* G. Velo and A.S. Wightman (Lecture Note in Physics, Vol. **73**, Springer, Berlin, 1976), *and references therein.*

[2] W. Rarita and J. Schwinger, Phys. Rev. **D60** (1941) 61.

[3] D.Z. Freedman, P. van Nieuwenhuizen and S. Ferrara, Phys. Rev. **D13** (1976) 3214;
P. van Nieuwenhuizen, Phys. Rep. **68C** (1981) 189.

[4] S. Deser and B. Zumino, Phys. Lett. **62B** (1976) 335.

[5] S.J. Gates, Jr., M.T. Grisaru, M. Roček and W. Siegel, *'Superspace'* Benjamin/Cummings, Reading, MA (1983).

[6] J. Wess and J. Bagger, *'Superspace and Supergravity'*, Princeton University Press (1992).

[7] R. Haag, J.T. Lopuszanski and M. Sohnius, Nucl. Phys. **B88** (1975) 257.

[8] W. Nahm, Nucl. Phys. **B135** (1978) 149.

[9] F.R. Ore, Jr. and P. van Nieuwenhuizen, Nucl. Phys. **B204** (1982) 317;
L. Baulieu, B. Grossman and R. Stora, Phys. Lett. **180B** (1986) 95;
A. Bouda, Phys. Rev. **D38** (1988) 3174.

[10] C. Becchi, A. Rouet and R. Stora, Ann. of Phys. **98** (1976) 287;
I.V. Tyutin, Lebedev-75-39 (1975).

[11] M. Henneaux, Phys. Rep. **126C** (1985) 1; L. Baulieu, *ibid.* **192C** (1985) 1.

[12] H. Nishino and S. Rajpoot, Phys. Rev. **D72** (2005) 085020; `hep-th/0508076`.

[13] N. Scherk and J.H. Schwarz, Nucl. Phys. **B153** (1979) 61.

[14] T. Kugo and P.K. Townsend, Nucl. Phys. **B211** (1983) 157.

[15] G. Velo and D. Zwanziger, Phys. Rev. **D186** (1969) 1337, *ibid.* **D188** (1969) 2218.

[16] K. Hayashi and T. Shirafuji, Prog. Theor. Phys. **64** (1980) 866; *ibid.* **64** (1980) 866; *ibid.* **64** (1980) 1435, Erratum-*ibid.* **66** (1981) 741; *ibid.* **64** (1980) 2222; *ibid.* **65** (1981) 525; *ibid.* **66** (1981) 318; *ibid.* **66** (1981) 2258.

[17] H. Nishino and S. Rajpoot, Phys. Rev. **D70** (2004) 027701, `hep-th/0402111`.

[18] H. Nishino and S. Rajpoot, Phys. Rev. **D71** (2005) 125002, `hep-th/0504097`.